\documentclass[sn-mathphys,Numbered]{sn-jnl}
\usepackage[sort&compress]{natbib}
\usepackage{graphicx}%
\usepackage{multirow}%
\usepackage{amsmath,amssymb,amsfonts}%
\usepackage{amsthm}%
\usepackage{mathrsfs}%
\usepackage[title]{appendix}%
\usepackage{xcolor}%
\usepackage{textcomp}%
\usepackage{manyfoot}%
\usepackage{booktabs}%
\usepackage{algorithm}%
\usepackage{algorithmicx}%
\usepackage{algpseudocode}%
\usepackage{listings}%
\usepackage{caption}
\usepackage{subcaption}
\usepackage{hyperref}
\hypersetup{
    colorlinks,
    citecolor=blue,
    filecolor=blue,
    linkcolor=blue,
    urlcolor=blue} 

\begin{document}

\title[Relaxation of relativistic four-component plasma]{Relaxation of a two electron-temperature relativistic hot electron-positron-ion plasma}

\author*{\fnm{Usman} \sur{Shazad*}}\email{usmangondle@gmail.com}

\author{\fnm{M.} \sur{Iqbal}}

\affil{\orgdiv{Department}, \orgname{University of Engineering and Technology}, \orgaddress{\city{Lahore}, \postcode{54890}, \country{Pakistan}}}

\abstract{The relaxation of a relativistic hot electron-positron plasma with a small fraction of cold electron and ion species is investigated, and a quadruple Beltrami (QB) relaxed state is derived. The QB state is a non-force-free state that is a linear combination of four single force-free states and is therefore comprised of four self-organized structures. Additionally, triple as well as double Beltrami states can be obtained by adjusting the generalized helicities of plasma species. The analysis of the relaxed states demonstrates that, for specific values of the generalized helicities, cold species density and relativistic temperature can transform paramagnetic structures into diamagnetic structures and vice versa. Also, when the flow vortices of plasma species are aligned with the magnetic field, it leads to a classically perfect diamagnetic relaxed state, and the strength of the magnetic field increases with a decrease in relativistic temperature and an increase in cold species density. The current investigation aims to enhance comprehension of energy transformation mechanisms and the formation of multiscale structures in astrophysical and laboratory plasmas.}


%%\pacs[JEL Classification]{D8, H51}

%%\pacs[MSC Classification]{35A01, 65L10, 65L12, 65L20, 65L70}

\maketitle

\section{Introduction} \label{S1}

The study of electron-positron-ion (EPI) plasmas continues to be a main focus of contemporary research related to the comprehension of the fundamental processes of the universe. Such plasmas are ubiquitous throughout the nature and can be found in a wide variety of astrophysical settings, such as early universe \cite{Misner1973,Rees1983}, pulsar magnetospheres  \cite{Michel1982,Pokhotelov2001}, an active galactic nuclei (AGN) \cite{Miller1987,Kirk1992}, the relativistic jet of a quasar \cite{Wardle1998}, the inner region of an accretion disc surrounding a black hole \cite{Liang1988}, solar atmosphere \cite{Hudson1995,Kozlovsky2004}, and etc. Besides these astrophysical environments the EPI plasmas can also be created in the laboratory through the process of relativistic heavy-ion collisions \cite{Baur1990}, the injection of positrons
into the electron ion system \cite{Greaves1997,Helander2003}, or the interaction of ultra-intense lasers with matter \cite{Sarri2015,Chen2023}. Most of the astrophysical and lab EPI plasmas mentioned above are made up of relativistically hot electron-positron (EP) pairs, but there are also a small number of heavy ions. The formation of various non-linear structures in relativistically hot EPI plasmas has been the subject of a significant amount of investigation over the course of the past few decades \cite{Berezhiani1994,Berezhiani1994a,Mahajan1998a,Tsintsadze2013,Rozina2014}. 

In astrophysical and laboratory plasmas, relaxation or magnetic self-organization is an important phenomena in addition to waves and instabilities. Magnetic self-organization is the process of minimizing the energy of the plasma under some helicity constraints. When the magnetic energy of plasma under magnetic helicity constraint is minimized, the Beltrami field (--a vector field in which the vortex is aligned with its field) describes the relaxed state of an ideal MHD plasma. This relaxed state is also known as the single Beltrami state or the Woltjer-Taylor state, and it can be represented as $\mathbf{\nabla}\times\mathbf{B}=\lambda \mathbf{B}$, where $\mathbf{B}$ is magnetic field and $\lambda$ is a constant and eigenvalue of the curl operator--measure of shear or twist in magnetic field. The eigenvalue $\lambda$ is also called the scale parameter and is dimensionally equal to the inverse of length; hence, the size and nature of relaxed state structure are determined by this constant. In a single Beltrami state, plasma is force-free and flowless because the current density is aligned with the magnetic field, and as a consequence, there is no pressure gradient in the plasma \cite{Woltjer1958,Taylor1974}. 

However, in two fluid plasmas, when the inertial effects of electrons are ignored, the relaxed state is the double Beltrami (DB) state. The DB state is a linear combination of two single Beltrami states and characterized by two scale parameters. The DB state is a non-force-free state that shows strong magnetofluid coupling and appreciable pressure gradients \cite{Mahajan1998,Steinhauer2002}. Whereas, when the inertial effects of both the plasma species are taken into account in two fluid plasmas, the self-organized state is the triple Beltrami (TB) state. The TB state is a linear superposition of three single force-free fields and is characterized by three scale parameters \cite{Bhattacharyya2003}. Furthermore, in a most recent study by Shatashvili et al., for a three-component plasma, when the inertia of all the plasma species is taken into account, the relaxed state is a quadruple Beltrami (QB) state. This QB state is the linear sum of four single Beltrami states and is characterized by four scale parameters \cite{Shatashvili2016}. In addition to strong magnetofluid coupling and pressure gradients, the creation of multiscale structures is another important feature that is present in the multi-Beltrami states.

During the course of the past few years, the study of plasma self-organization has expanded to relativistic plasmas.
For instance, a study by Iqbal et al. shows that a relativistic hot EP plasma self-organizes into a TB state \cite{Iqbal2008}. A plasma is said to be relativistically hot when the thermal energy of its species is equal to or greater than their rest mass energy. Likewise, the relaxed state of a relativistically hot EPI plasma is also a TB state \cite{Iqbal2012,Iqbal2013,Usman2021,Usman2023a}. In the most recent study by Shazad and Iqbal, it is demonstrated that for three distinct Beltrami parameters, the relaxed state of relativistic hot EPI is a QB state \cite{Usman2023b}. The relaxed states in relativistic degenerate plasmas have been focus of Shatashvili and co-workers. It has been investigated that a relaxed state of relativistic dense degenerate EPI plasma is the QB state \cite{Shatashvili2016}. Similarly, the relaxed state of three-component electron-ion plasma, composed of relativistic degenerate and classical relativistic hot electrons and classical ions, is also a QB state \cite{Shatashvili2019}. For the plasmas observed in the vicinity of black holes, the Beltrami states are also of significant interest; in such plasmas, the implications of general relativity are taken into account and investigated by many researchers \cite{Bhattacharjee2015,Bhattacharjee2019,Bhattacharjee2020,Asenjo2019,Bhattacharjee2020a}. Most recently, Bhattacharjee  \cite{Bhattacharjee2023} and Usman and Iqbal  \cite{Usman2023c} have also explored the relaxed states of single and two fluid plasmas in a massive photon field.
Their approach involved the utilization of Proca electrodynamics, which allowed for the incorporation of the influence of non-zero photon mass. It is also important to emphasize that the above-mentioned multi-Beltrami relaxed states are applied to study a wide range of physical phenomena that occur in the laboratory and in space plasmas, such as high beta relaxed states in tokamak \cite{Mahajan2000,Yoshida2001}, eruptive events in solar plasma \cite{Ohsaki2002,Kagan2010}, flow generation \cite{Mahajan2002,Barnaveli2017}, striped wind phenomena in the pulsar nebula \cite{Pino2010}, dynamo and reverse dynamo mechanisms \cite%
{Mininni2002,Mahajan2005,Lingam2015,Kotorashvili2020,Kotorashvili2022}.

The aim of the present work is to investigate the relaxation of two electron-temperature
relativistic hot EPI plasmas. It is well established that cold electron and heavy ion species can coexist with hot electron–positron pairs in astrophysical plasmas. For instance, when the outflows of hot electron-positron pair plasma from pulsars encounter low-density electron-ion plasma in the interstellar medium, a two electron-temperature relativistic hot EPI plasma is created.
In such plasmas within the framework of pulsars and AGN, non-linear wave dynamics and the formation of solitons have been studied in Refs. \cite{Berezhiani1995,Shatashvili1997,Shatashvili1999}.
Most recently, Dieckmann and co-workers have extensively investigated the expansion of pair clouds into electron-proton plasmas, ion acoustic solitary waves, filamentation instability, Weibel instability, collisionless Rayleigh-Taylor-like instability, cocoon formation, and other plasma phenomena in such plasmas \cite{Dieckmann2018,Dieckmann2018a,Dieckmann2018b,Dieckmann2019,Dieckmann2020,Dieckmann2020a,Dieckmann2022}.

In addition to the nonlinear processes described above, equilibrium or relaxed states of two electron temperature relativistic hot EPI plasma (--composed of hot electrons and positrons as well as cold electrons and heavy positively charged ions) have not been investigated. In order to derive a relaxed state equation, the inertia of hot electrons, positrons, and cold electrons has been accounted for, whereas ions have been assumed to be static. Three Beltrami conditions result from the steady-state solution of vortex dynamics equations. By combining these Beltrami conditions with Ampere's law, a QB relaxed state is obtained. The investigation of the QB state demonstrates that the density of cold plasma species and relativistic temperature for certain values of Beltrami parameters can transform diamagnetic structures into paramagnetic ones. Furthermore, it is also shown that by adjusting the generalized helicities of plasma species, TB, DB, and classical perfect diamagnetic states can also be obtained, and the impact of plasma parameters on these lower-index Beltrami states is also illustrated.

The structure of the paper is as follows: In Sec. \ref{S2}, the QB state is derived from macroscopic evolution equations. The characteristics of scale parameters are described in Sec. \ref{S3}. In Sec. \ref{S4}, an analytical solution for the QB state in a simple slab geometry is given, and the impact of plasma parameters on the relaxed state structures is studied. By adjusting the generalized helicities of plasma species, the lower index Beltrami states are derived in Sec. \ref{S5}. The summary and conclusion of the present study are given in Sec. \ref{S6}.

\section{Model Equations and QB state}\label{S2}

Consider a quasineutral, magnetized, four-component plasma composed of
relativistically hot electrons and positrons with a small fraction of cold
electrons (electrons are warm but assumed to be cold as compared to hot
electrons) and static positively charged ions. The quasineutrality condition
for this plasma system can be expressed as
\begin{equation}
N_{i}+N_{p}-N_{c}=1,  \label{QN}
\end{equation}
where $N_{i}=n_{i}/n_{h}$, $N_{p}=n_{p}/n_{h}$ and $N_{c}=n_{c}/n_{h}$, in
which $n_{i}$, $n_{p}$, $n_{h}$ and $n_{c}$ are ion, hot positrons, hot
electrons and cold electrons number densities respectively. Before looking
at the equations of motion, it's worth noting that EPI plasma has two
distinct kinds of relativistic regimes. In space and astrophysical settings,
the electromagnetic radiation emitted by luminous objects acts as a source
of powerful electromagnetic fields, allowing plasma particles to approach
relativistic quiver velocities. Hence, the masses of the particles become
functions of their respective velocities, whereas the relativistic mass
variation produces a multitude of significant physical effects. On the other
hand, at extremely high temperatures, the thermal energy of plasma particles
is equal to or greater than the energy at rest; this is a form of
relativistic regime that may have been particularly significant in the early
epochs of the universe. In the following, we will assume that the velocity
distribution of plasma particles is local relativistic~Maxwellian, then by
following Ref. \cite{Berezhiani1995}, the relativistic equation of motion
for $\alpha$ plasma species ($\alpha$=$h$, $p$ and $c$--represents hot electron, positron and cold electron species, respectively) can be expressed as:
\begin{equation}
\frac{\partial \mathbf{\Pi}_{\alpha }}{\partial t}+\frac{1}{n_{\alpha }}
\mathbf{\nabla }\mathit{p}_{\alpha }=q_{\alpha }\mathbf{E}+\mathbf{V}
_{\alpha }\times \left( \mathbf{\Pi}_{\alpha }+\frac{q_{\alpha }}{c}\mathbf{B}
\right) ,  \label{REM}
\end{equation}
where  $\mathbf{\Pi}_{\alpha }=\gamma _{\alpha }m_{0\alpha }G_{\alpha }\mathbf{V%
}_{\alpha }$, $\mathbf{V}_{\alpha }$, $m_{0\alpha }$, $q_{\alpha }$, $%
\mathit{p}_{\alpha }=\left( n_{\alpha }/\gamma _{\alpha }\right) T_{\alpha }$
, $n_{\alpha }$, $T_{\alpha }$ and $\gamma _{\alpha }=\left( 1-V_{\alpha
}^{2}/c^{2}\right) ^{-1/2}$ are relativistic momentum, velocity, rest mass,
electric charge, relativistic pressure, number density, proper temperature
and relativistic factor of $\alpha $ plasma species, respectively whereas $c$
is speed of light, $\mathbf{E}=\mathbf{-\nabla }\phi -c^{-1}\partial \mathbf{%
A/\partial }t$ and $\mathbf{B}=\mathbf{\nabla }\times \mathbf{A}$ are
electric and magnetic fields that are related to scalar and vector
potentials $\phi $ and $\mathbf{A}$, respectively. In the equation of motion
(\ref{REM}), the thermal relativistic effects appear through the factor $%
G_{\alpha }=K_{3}(z_{\alpha })/K_{2}(z_{\alpha })$---say relativistic
temperature, where $K_{\nu }$ are the modified Bessel functions and $%
z_{\alpha }=m_{0\alpha }c^{2}/T_{\alpha }$. For non-relativistic case $%
T_{\alpha }<<m_{0\alpha }c^{2}$, and the factor $G_{\alpha }$ can
approximately be taken as $G_{\alpha }\simeq 1+5T_{\alpha }/2m_{0\alpha
}c^{2}$, whereas for ultra- relativistic regime, $T_{\alpha }>>m_{0\alpha
}c^{2}$ and $G_{\alpha }$ can be approximated as $G_{\alpha }\simeq
4T_{\alpha }/m_{0\alpha }c^{2}$. It is essential to point out that the
equation (\ref{REM}) is augmented with the following equation of state: 
\begin{equation}
\frac{n_{\alpha }}{\gamma _{\alpha }}\frac{z_{\alpha }}{K_{2}\left(
z_{\alpha }\right) }\exp \left( -z_{\alpha }G_{\alpha }\right) =\text{
constant.}  \label{ES}
\end{equation}
To make our model simplified and more suitable to analytical studies we
assume that the fluid velocity of plasma species is non-relativistic ($%
\gamma _{\alpha }\approx 1$), and the hot electron and positron temperatures
are same and constant ($G_{h}=G_{p}=G$) while for cold electron species $%
G_{c}=1$. To normalize plasma species velocities, magnetic field, length and
time in the model equations, we will use Alfv\'{e}n speed $v_{A}=B_{0}/%
\sqrt{4\pi m_{0}n_{h}}$, some arbitrary value of magnetic field $B_{0}$,
electron skin depth $l_{e}=\sqrt{m_{0}c^{2}/4\pi n_{h}e^{2}}$ and the
inverse of plasma frequency $\omega _{pe}^{-1}=l_{e}/c$, respectively, where 
$m_{0}$ and $e$ are electron rest mass and elementary charge. Then, the equations
of motions for dynamic plasma species in normalized form are as follows 
\begin{equation}
\frac{\partial \mathbf{P}_{h}}{\partial t} =\mathbf{V}_{h}\times \mathbf{\Omega }_{h}
-\mathbf{\nabla }\psi _{h},  \label{EME}
\end{equation}
\begin{equation}
\frac{\partial \mathbf{P}_{p}}{\partial t} =\mathbf{V}_{p}\times \mathbf{\Omega }_{p}
-\mathbf{\nabla }\psi _{p},  \label{EMP}
\end{equation}
\begin{equation}
\frac{\partial \mathbf{P}_{c}}{\partial t} =\mathbf{V}_{c}\times \mathbf{\Omega }_{c}
-\mathbf{\nabla }\psi _{c},  \label{EMC}
\end{equation}
where $\mathbf{V}_{h}$, $\mathbf{V}_{p}$ and $\mathbf{V}_{c}$ are velocities
of hot electron, hot positron and cold electron species, $\psi _{h}=-\phi -\mu G$, $\psi
_{p}=\phi -\mu G$ and $\psi _{c}=-\phi +V_{c}^{2}/2+p_{e}/N_{c}$
,  $p_{e}$ is cold electron pressure
and $\mu=c^{2}/v_{A}^{2}$.  In equations (\ref{EME}-\ref{EMC}), $\mathbf{P}%
_{h}=G\mathbf{V}_{h}-\mathbf{A}$, $\mathbf{P}_{p}=G\mathbf{V}_{p}-\mathbf{B}$
and $\mathbf{P}_{c}=\mathbf{V}_{c}-\mathbf{A}$ are the generalized or canonical
momentum of plasma species. Whereas, the curl of the generalized or canonical
momentum is called generalized or canonical vorticity ($\mathbf{\Omega }%
_{\alpha }=$\textbf{$\nabla $}$\times \mathbf{P}_{\alpha }$) of plasma
species. Now, by taking curl of the macroscopic evolution equations (\ref{EME}-\ref{EMC}), one can
obtain the following vortex dynamics equations
\begin{equation}
\frac{\partial \mathbf{\Omega }_{h}}{\partial t}=\mathbf{\nabla }\times
\left( \mathbf{V}_{h}\times \mathbf{\Omega }_{h}\right) ,  \label{VEE}
\end{equation}
\begin{equation}
\frac{\partial \mathbf{\Omega }_{p}}{\partial t}=\mathbf{\nabla }\times
\left( \mathbf{V}_{p}\times \mathbf{\Omega }_{p}\right) ,  \label{VEP}
\end{equation}
\begin{equation}
\frac{\partial \mathbf{\Omega }_{c}}{\partial t}=\mathbf{\nabla }\times
\left( \mathbf{V}_{c}\times \mathbf{\Omega }_{c}\right).
\label{VEC}
\end{equation}
From above equations also note that $\psi _{h,p,c}$ includes all possible
variables which do not directly influence the evolution of vorticity of
plasma species. The following Beltrami conditions result from the steady state
solution of vorticity evolution equations  (\ref{VEE}-\ref{VEC}) when all the gradient forces are restricted to zero ($%
\mathbf{\nabla }\psi _{h,p,c}=0$): 
\begin{equation}
\mathbf{\nabla }\times G\mathbf{V}_{h}-\mathbf{B}=a_{h}G\mathbf{V}_{h},
\label{BEE}
\end{equation}
\begin{equation}
\mathbf{\nabla }\times G\mathbf{V}_{p}+\mathbf{B}=a_{p}G\mathbf{V}_{p},
\label{BEP}
\end{equation}
\begin{equation}
\mathbf{\nabla }\times \mathbf{V}_{c}-\mathbf{B}=a_{c}\mathbf{V}_{c}.  \label{BEC}
\end{equation}
where $a_{h}$, $a_{p}$ and $a_{c}$ are Beltrami parameters. Beltrami
parameters are the measure of twist and shear in magnetic field and flow and are linked
with the generalized helicities of the plasma species that are the constants of motion. By following Ref. 
\cite{Steinhauer1997}, from equations of motion, vortex dynamics equations
and Ampere's law we can show that the magnetofluid energy ($E_{mf}$) and
generalized helicities for hot electrons ($h_{h}$), positrons ($h_{p}$) and
cold electrons ($h_{c}$) are constants of motion for this plasma model, and
these constants of motion can be expressed in the following manner: $%
E_{mf}=\left\langle
GV_{e}^{2}+GN_{p}V_{p}^{2}+N_{c}V_{c}^{2}+B^{2}\right\rangle $, $%
h_{h}=\left\langle \mathbf{P}_{h}\cdot \mathbf{\Omega }_{h}\right\rangle $, $%
h_{p}=\left\langle \mathbf{P}_{p}\cdot \mathbf{\Omega }_{p}\right\rangle $
and $h_{c}=\left\langle \mathbf{P}_{c}\cdot \mathbf{\Omega }%
_{c}\right\rangle $, where $\left\langle ..\right\rangle $ represents the
volume integral. These constants of motion also show that there are $n+1$
constants of motion ($E_{mf}$, $h_{h}$, $h_{p}$ and $h_{c}$) for $n$
---dynamic plasma species \cite{Mahajan2015}. Now, in order to couple the
independent dynamics of plasma species, we use Ampere's law under the
premise that the displacement current is negligible, that is given by
\begin{equation}
\mathbf{\nabla }\times \mathbf{B}=N_{p}\mathbf{V}_{p}-N_{c}\mathbf{V}_{c}-
\mathbf{V}_{h}.  \label{AL}
\end{equation}
To derive a relaxed state equation, we simultaneously solve equations (\ref%
{BEE}-\ref{AL}). From these equations, cold electron species velocity in terms of magnetic field is
\begin{equation}
\mathbf{V}_{c}=e_{4}\left(\mathbf{\nabla }\times\right)^{3}\mathbf{B}%
-e_{3}\left(\mathbf{\nabla }\times\right)^{2}\mathbf{B}+e_{2}\mathbf{%
\nabla }\times \mathbf{B}-e_{1}\mathbf{B},  \label{VC}
\end{equation}
where $(\mathbf{\nabla }\times)^{3}=\mathbf{\nabla }\times\mathbf{\nabla }\times\mathbf{\nabla }\times$, $(\mathbf{\nabla }\times)^{2}=\mathbf{\nabla }\times\mathbf{\nabla }\times$, $\alpha _{1}=\left( 1+N_{p}+GN_{c}\right) G^{-1}$, $\alpha _{2}=\alpha
_{1}a_{p}-N_{c}\left( a_{c}-a_{h}\right) -G^{-1}N_{p}\left(
a_{p}-a_{h}\right) $, $e_{1}=\alpha
_{2}e_{4}$, $e_{2}=\left( \alpha _{1}+a_{h}a_{p}\right) e_{4}$, $e_{3}=\left( a_{h}+a_{p}\right)
e_{4}$ and $e_{4}=\left[ \left( a_{c}-a_{h}\right) \left(
a_{p}-a_{c}\right) N_{c}\right] ^{-1}$. Similarly the expression for hot positron species velocity is
\begin{equation}
\mathbf{V}_{p}=p_{4}\left(\mathbf{\nabla }\times\right)^{3}\mathbf{B}
-p_{3}\left(\mathbf{\nabla }\times\right)^{2}\mathbf{B}+p_{2}\mathbf{
\nabla }\times \mathbf{B}-p_{1}\mathbf{B}.  \label{VP}
\end{equation}
where $p_{1}=\left( \alpha _{2}+\alpha
_{1}\left( a_{c}-a_{p}\right) \right) p_{4}$, $p_{2}=\left(
\alpha _{1}+a_{c}a_{h}\right) p_{4}$, $p_{3}=\left( a_{h}+a_{c}\right) p_{4}$  and $p_{4}=( a_{p}-a_{h})^{-1} ( a_{p}-a_{c})^{-1} N_{p}%
 ^{-1}$. By substituting the values of cold electron and hot positron species velocities in equation (\ref{AL}), we obtain the following value of hot electron species velocity  
\begin{equation}
\mathbf{V}_{h}=h_{4}\left(\mathbf{\nabla }\times\right)^{3}\mathbf{B}%
-h_{3}\left(\mathbf{\nabla }\times\right)^{2}\mathbf{B}+h_{2}\mathbf{
\nabla }\times \mathbf{B}-h_{1}\mathbf{B}  \label{VH}
\end{equation}
where $h_{1}=\left( \alpha _{2}+\alpha
_{1}\left( a_{c}-a_{h}\right) \right) h_{4}$, $h_{2}=\left(
\alpha _{1}+a_{c}a_{p}\right) h_{4}$, $h_{3}=\left( a_{c}+a_{p}\right) h_{4}$ and $h_{4}=( a_{p}-a_{h})^{-1}( a_{h}-a_{c})^{-1}$. By putting the value of $\textbf{V}_{h}$ in equation (\ref{BEE}) , we get the following QB relaxed state equation in terms of magnetic field
\begin{equation}
\left(\mathbf{\nabla }\times\right)^{4}\mathbf{B}-k_{4}\left(\mathbf{\nabla }\times\right)^{3}\mathbf{B}+k_{3}\left(\mathbf{\nabla }\times\right)^{2}\mathbf{B}-k_{2}\mathbf{\nabla \times B}+k_{1}\mathbf{B}=0,
\label{QB}
\end{equation}
where $(\mathbf{\nabla }\times)^{4}=\mathbf{\nabla }\times\mathbf{\nabla }\times\mathbf{\nabla }\times\mathbf{\nabla }\times$, $k_{4}=a_{h}+a_{p}+a_{c}$, $k_{3}=a_{h}a_{p}+a_{h}a_{c}+a_{p}a_{c}+
\left( 1+N_{p}+GN_{c}\right) G^{-1}$, $k_{2}=a_{c}a_{h}a_{p}+N_{c}\left(
a_{h}+a_{p}\right) +\left( a_{c}+a_{p}+N_{p}\left( a_{c}+a_{h}\right)
\right) G^{-1}$ and $k_{1}=a_{c}G^{-1}\left( a_{p}+N_{p}a_{h}\right)
+a_{h}a_{p}N_{c}$. The above QB equation (\ref{QB}) is the result of taking
into account the inertia of hot electrons, positrons and cold electrons, as
well as distinct Beltrami parameters for each plasma species whereas positively charged ions are assumed static and play their role through quasineutrality condition. Furthermore,  all the flow fields ($\textbf{V}_{h}$, $\textbf{V}_{p}$ and $\textbf{V}_{c}$) are also QB fields and it is an indication of strong magnetofluid coupling. It is worth noting that in this plasma model, if the ion species are also presumed to be dynamical in this plasma model, then there will be four Beltrami conditions, and a QB state can be obtained for four distinct Beltrami parameters if the inertia of cold electron species is neglected.  Alternatively, if all plasma species are presumed to be inertial and only three distinct Beltrami parameters are utilized, then a QB can also be derived.

 As mentioned earlier, the existence of cold plasma species in the hot pair plasma of pulsar magnetospheres is well established. Additionally, a number of researchers have investigated hot-pair plasmas with some cold plasma species, and they have found that cold species density also plays a crucial role in the formation of a wide range of non-linear structures \cite{Bharuthram1992,Pillay1992,Verheest1996,Lazarus2008,Lazarus2012}. Hence, for the sake of our analysis of the relaxed state, which will be presented in the following sections, we will adopt the typical plasma parameters of the pulsar magnetosphere, in which the number density of hot electron or positron species is  $n_{h}=n_{p}=10^{6}$ cm$^{-3}$ at a distance of 100 radii  from the surface of the pulsar, and the skin depth is $l_{e}=5.4\times10^{2}$cm \cite{Melrose1978}. Since $n_{h}=n_{p}$, we shall refer to $N_{p}=1.0$ and $N_{i}=N_{c}$ throughout this study. Importantly, the values of other plasma parameters such as cold species density, relativistic temperature, Beltrami parameters and distances used in plotting field profiles are all arbitrary.  So, the objective of this study is to demonstrate that by manipulating these plasma parameters, the size and nature of the relaxed state structures can be transformed, which can aid in the comprehension of energy transformation mechanisms, eruptive events, unified dynamo/reverse dynamo mechanisms, and other related plasma phenomena.
\section{Characteristics of scale-parameters} \label{S3}

\begin{figure}[h!]
\centering
\includegraphics[scale=0.9]{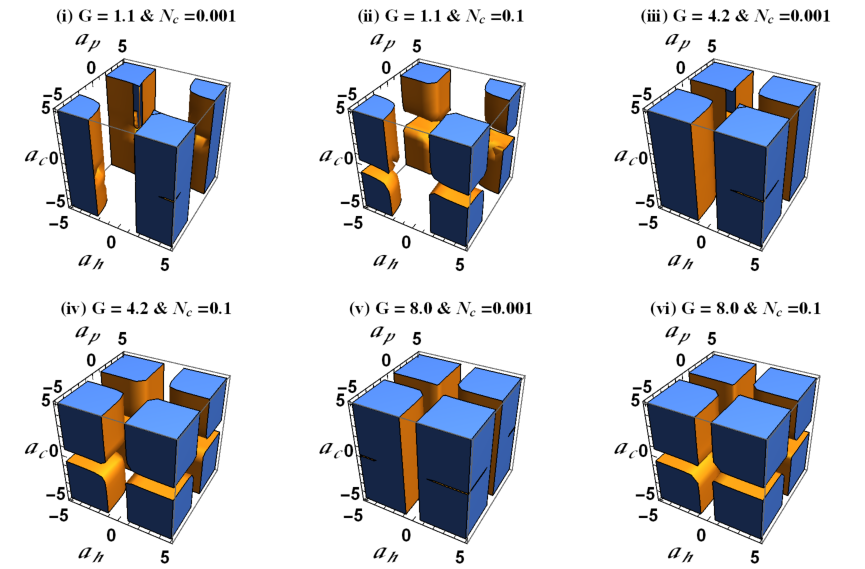}
\caption{The plot of $D>0$ as a function of Beltrami parameters ($a_{h}$, $a_{p}$ and $a_{c}$) depicting the nature of scale parameters for different values of $N_{c}$ and $G$. In the colored region all the scale parameters are real.}
\label{fig:1}
\end{figure}

As the curl operators are commutative in nature which enables us to express the QB
equation (\ref{QB}) as the superposition of four distinct Beltrami fields $%
\mathbf{B}_{j}$ ($j=1,2,3,4$). These Beltrami fields fulfill the following
condition $\mathbf{\nabla }\times \mathbf{B}_{j}=\lambda _{j}\mathbf{B}_{j}$
, where $\lambda _{j}$ and $\mathbf{B}_{j}$ are eigenvalues and
eigenfunctions of the curl operator respectively \cite{Yoshida1990}.
Introducing the concept of eigenvalues, it is possible to express equation (
\ref{QB}) in the following manner 
\begin{equation}
\left( \text{curl}-\lambda _{1}\right) \left( \text{curl}-\lambda
_{2}\right) \left( \text{curl}-\lambda _{3}\right) \left( \text{curl}
-\lambda _{4}\right) \mathbf{B}=0.  \label{SQB}
\end{equation}
The relationship between these scale parameters and plasma parameters from
equations (\ref{QB} and \ref{SQB}) is as follows: 
\begin{eqnarray*}
k_{1} &=&\lambda _{1}\lambda _{2}\lambda _{3}\lambda _{4}, \\
k_{2} &=&\lambda _{1}\lambda _{2}\lambda _{3}+\lambda _{1}\lambda
_{2}\lambda _{4}+\lambda _{2}\lambda _{3}\lambda _{4}+\lambda _{1}\lambda
_{3}\lambda _{4}, \\
k_{3} &=&\lambda _{1}\lambda _{2}+\lambda _{2}\lambda _{3}+\lambda
_{1}\lambda _{3}+\lambda _{1}\lambda _{4}+\lambda _{2}\lambda _{4}+\lambda
_{3}\lambda _{4}, \\
k_{4} &=&\lambda _{1}+\lambda _{2}+\lambda _{3}+\lambda _{4}.
\end{eqnarray*}
The above relationship between these scale parameters and coefficients of QB
equation ( \ref{QB}) satisfy the Vieta's formulas which dictates us to write the following quartic equation 
\begin{equation}
\lambda ^{4}-k_{4}\lambda ^{3}+k_{3}\lambda ^{2}-k_{2}\lambda +k_{1}=0,
\label{QEVE}
\end{equation}
 and the roots of equation (\ref{QEVE}) give us the values of four scale
parameters $\lambda _{1}$, $\lambda _{2}$, $\lambda _{3}$, and $\lambda _{4}$. In the QB
state, the scale parameters may be either real or both real and complex,
depending on the plasma parameters. To determine the nature of the scale
parameters of QB state, the discriminant ($D$) of a quartic equation (\ref%
{QEVE}) can be used. When $D>0$, all the eigenvalues are real, whereas for $%
D<0$, two of the eigenvalues are real and the other two are complex
conjugate pair. The plot of $D>0,$ as a function of Beltrami parameters ($%
a_{h},$ $a_{p}$ and $a_{c}$) for different values of cold electron species
density ($N_{c}$) and relativistic temperature ($G$) is presented in figure (\ref{fig:1}), and it offers a glimpse of the impact that plasma parameters
have on the nature of scale parameters. The colored region in figure (\ref
{fig:1}) indicates the existence of all four real eigenvalues, whereas in
the transparent region, two of the eigenvalues are real and the other two
are complex conjugate pair. From the plots in figure (\ref{fig:1}), it is
evident that for higher $N_{c}$, some of the real roots become complex,
whereas for higher $G$, some of the complex eigenvalues become real. As the
real scale parameters allow for the formation of paramagnetic structures in
a relaxed state, however the combination of real and complex scale
parameters leads to diamagnetic or partial diamagnetic structures. So, from
figure (\ref{fig:1}), it is clear that for different plasma settings both
para- and diamagnetic structures can be created.
\begin{figure}[h!]
     \centering
     \begin{subfigure}[b]{0.49\textwidth}
         \centering
         \includegraphics[width=\textwidth]{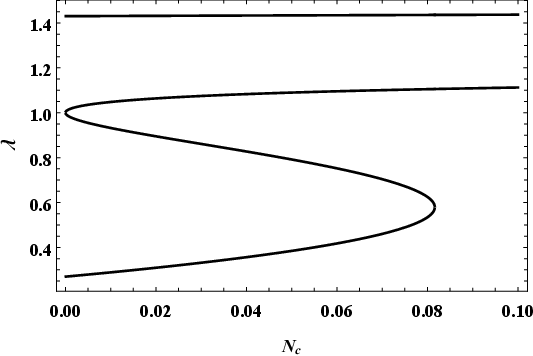}
         \caption{$a_{h}=1.2$, $a_{p}=1.5$, $a_{c}=1.0$ and $G=7.0$.}
         \label{fig:2a}
     \end{subfigure}
     \hfill
     \begin{subfigure}[b]{0.49\textwidth}
         \centering
         \includegraphics[width=\textwidth]{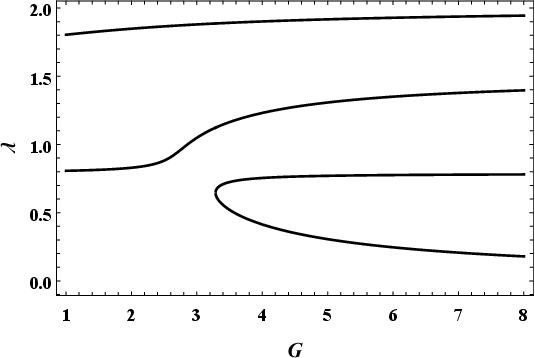}
         \caption{$a_{h}=2.0$, $a_{p}=1.5$, $a_{c}=0.8$ and $N_{c}=0.01$.}
         \label{fig:2b}
     \end{subfigure}
        \caption{The nature and variation of scale parameters ($\lambda_{j}$) as a function of $N_{c}$ and $G$ for given values of plasma parameters.}
        \label{fig:2}
\end{figure}

Next, we explore the impact of $N_{c}$ and $G$ on the nature and variation
of scale parameters for the fixed values of Beltrami parameters. In this
context, in figure (\ref{fig:2}), scale parameters are plotted as a function
of $N_{c}$ and $G$ for given Beltrami parameters. For $a_{h}=1\mathbf{.}2$, $%
a_{p}=1\mathbf{.}5$, $a_{c}=1\mathbf{.}0$ and $G=7\mathbf{.}0$, the figure (%
\ref{fig:2a}), shows the nature and variation in eigenvalues of QB state as
a function of $N_{c}$. The plot demonstrates that at lower densities, all
of the scale parameters are real and distinct; however, as density
increases, two of the scale parameters become complex, while the other two
remain real and distinct. Second, when density drops, the values of two
scale parameters vary (one of them decreases while the other increases),
while the values of the other two scale parameters remain constant. Similarly, the
effect of relativistic temperature on the variation of the sizes and nature
of self-organized vortices is portrayed in figure (\ref{fig:2b}). For this
purpose, the plasma parameters $a=2\mathbf{.}0$, $b=1\mathbf{.}5$, $c=0%
\mathbf{.}8$ and $N_{c}=0\mathbf{.}01$ are used. Clearly, at lower
relativistic temperatures, two scale parameters are real and the other two
are complex, but as the relativistic temperature increases, all of the scale
parameters are real. It is clear from the plot that only the small scale
parameter is primarily influenced by higher relativistic temperatures, and
it gets shorter and shorter as thermal energy increases while the other
three eigenvalues almost remain constant.
\section{Analytical solution of QB state} \label{S4}

\begin{figure}
     \centering
     \begin{subfigure}[b]{0.49\textwidth}
         \centering
         \includegraphics[width=\textwidth]{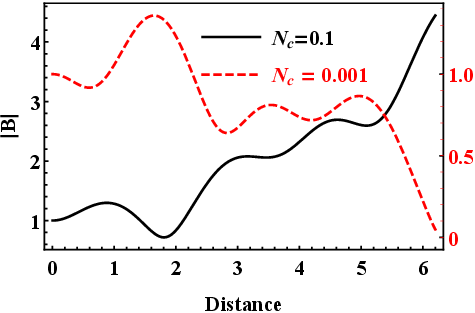}
         \caption{$a_{h}=2.4$, $a_{p}=4.0$, $a_{c}=0.5$, $G=8.0$ and $N_{c}=0.1$--(left vertical axis) and $0.001$--(right vertical axis)}
         \label{fig:3a}
     \end{subfigure}
     \hfill
     \begin{subfigure}[b]{0.49\textwidth}
         \centering
         \includegraphics[width=\textwidth]{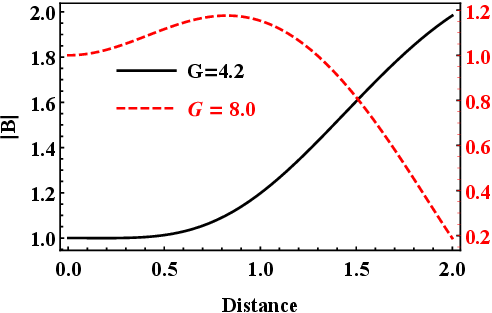}
         \caption{$a_{h}=2.0$, $a_{p}=1.0$, $a_{c}=3.5$, $N_{c}=0.01$ and $G=4.2$--(left vertical axis) and $8.0$--(right vertical axis) .}
         \label{fig:3b}
     \end{subfigure}
        \caption{Variation in QB magnetic structures for different values of plasma parameters.}
        \label{fig:3}
\end{figure}

We will employ a simple slab geometry for the analytical solution of the QB state to demonstrate the impact of
plasma parameters on the formation and nature of  magnetic
structures. The general solution to QB equation (\ref{QB}) may be expressed
as the linear combination of four distinct Beltrami fields, and it can be
written as $B=\sum\limits_{j=1}^{4}C_{j}B_{j}$, where $C_{j}$ are constants
and their values can be calculated by using suitable boundary conditions.
So, the analytical solution of QB equation (\ref{QB}) in a simple slab geometry is given by 
\begin{equation}
\mathbf{B=}\sum\limits_{j=1}^{4}C_{j}\left[ \sin \left( \lambda _{j}x\right) 
\widehat{y}+\cos \left( \lambda _{j}x\right) \widehat{z}\right],
\end{equation}
 where $C_{j}$'s are constants and their values can be obtained with the help of following boundary
conditions: $\left\vert \mathbf{B}_{z}\right\vert
_{x=0}=\sum\limits_{j=1}^{4}C_{j}=b_{1}$, $\left\vert \mathbf{B}%
_{y}\right\vert _{x=x_{0}}=\sum\limits_{j=1}^{4}C_{j}\sin \left( \lambda
_{j}x_{0}\right) =b_{2}$, $\left\vert (\mathbf{\nabla \times B}%
)_{z}\right\vert _{x=0}=\sum\limits_{j=1}^{4}C_{j}\lambda _{j}=b_{3}$ and$%
\left\vert (\mathbf{\nabla }\times \mathbf{B)}_{y}\right\vert
_{x=x_{0}}=\sum\limits_{j=1}^{4}C_{j}\lambda _{j}\sin \left( \lambda
_{j}x_{0}\right) =b_{4}$, where $b_{1}$, $b_{2}$, $b_{3}$, $b_{4}$, and $x_{0}
$ are arbitrary and real valued constants. Now, by using certain values of the Beltrami parameters and
boundary conditions ($b_{1}=1.0$, $b_{2}=0.25$, $b_{3}=0.4$, and $b_{4}=0.5$), we will demonstrate that the cold species density and relativistic temperature
can transform paramagnetic structures into diamagnetic
structures and vice versa. Figure (\ref{fig:3a}) illustrates the impact of cold species density on the magnetic field
structures for the fixed values of Beltrami parameters and relativistic temperature $a_{h}=2.4$, $a_{p}=4.0$, $a_{c}=0.5$ and $G=8.0$.  It shows that for $N_{c}=0.001$, the scale parameters are real and given by $\lambda
_{1}=0.0885$, $\lambda _{2}=0.4975$, $\lambda _{3}=2.3449$ and $\lambda_{4}=3.9691$. As all of the
scale parameters are real, figure (\ref{fig:3a}) depicts the formation of a
paramagnetic structure. However, for $
N_{c}=0\mathbf{.}1$, the eigenvalues are $\lambda _{1}=2.432$, $\lambda _{2}=3.9694$  and $\lambda
_{3,4}=0.2922\pm 0\mathbf{.}2463i$. Since the scale parameters are combination of real and complex values,
and consequently the self-organized magnetic structures show
diamagnetic behavior.

In order to demonstrate the effect of relativistic temperature on the
magnetic field structures, we take the following values of the plasma
parameters $a_{h}=2.0$, $a_{p}=1.0$, $a_{c}=3.5$ and $N_{c}=0.01$. In case of moderately relativistic regime, for instance, when $G$ is $4
\mathbf{.}2$, then two scale parameters are real ($\lambda _{1}=1\mathbf{.}8893$, $\lambda _{2}=3\mathbf{.}4973$) while
other two are complex conjugate ($\lambda _{3,4}=0\mathbf{.}5566\pm 1\mathbf{.}2674i$). For these real and complex eigenvalues figure (\ref{fig:3b}) depicts the diamagnetic trend. Contrary to this, for ultra-relativistic regime, consider $G=8\mathbf{.}0$,
then the scale parameters change their nature and all the scale parameters
become real. The values of the scale parameters are $\lambda _{1}=0\mathbf{.}2380$, $
\lambda _{2}=0\mathbf{.}8252$, $\lambda _{3}=1\mathbf{.}9395$ and $\lambda
_{4}=3\mathbf{.}4973$. For these real scale parameters, the relaxed state shows paramagnetic trend. 

From the above discussion, it can be concluded that for suitable Beltrami parameters and boundary conditions, the cold species density and relativistic temperature of hot pair species have a key role in controlling the nature of the relaxed state structures. At higher relativistic temperatures and lower cold species densities, magnetic structures are paramagnetic, while at lower relativistic temperatures and higher cold species densities, diamagnetic structures are formed for some specific values of Beltrami parameters along with suitable boundary conditions.

\section{Role of generalized helicities and lower index Beltrami states} \label{S5}

In order to derive the QB state, we have used three distinct Beltrami parameters for each plasma species. As a result, this state is a more generalized version of the relaxed state. As was previously mentioned, the Beltrami parameters are the measure of the generalized helicities of the plasma species, and linked with the constants of motion for the plasma system. So, by manipulating the Beltrami parameters or generalized helicities of the plasma species, one is able to obtain the lower index Beltrami states such as TB and DB as well as the classical perfect diamagnetic state. To demonstrate this, we consider some special cases.

\subsection{TB state} \label{ss4.1}

\begin{figure}
     \centering
     \begin{subfigure}[b]{0.49\textwidth}
         \centering
         \includegraphics[width=\textwidth]{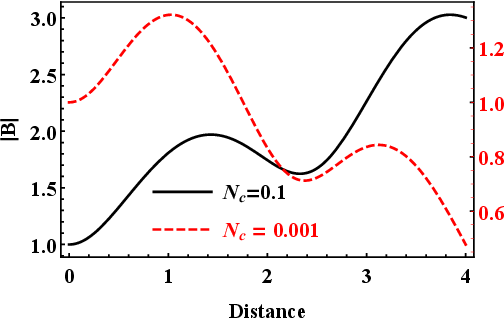}
         \caption{$a=3.5$, $a_{c}=0.85$, $G=4.2$ and $N_{c}=0.1$--(left vertical axis) and $0.001$--(right vertical axis).}
         \label{fig:4a}
     \end{subfigure}
     \hfill
     \begin{subfigure}[b]{0.49\textwidth}
         \centering
         \includegraphics[width=\textwidth]{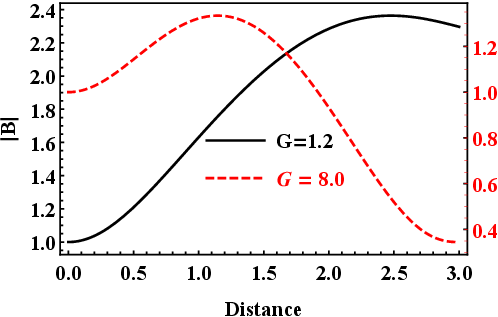}
         \caption{$a=0.9$, $a_{c}=3.0$, $N_{c}=0.01$ and $G=1.2$--(left vertical axis) and $8.0$--(right vertical axis).}
         \label{fig:4b}
     \end{subfigure}
        \caption{Trend of TB magnetic structures for different values of plasma parameters.}
        \label{fig:4}
\end{figure}

Consider the case when the generalized helicities of hot pair species are equal, such that the ratios of generalized vorticities to the respective flows are equal. So for this particular case, $a_{h}=a_{p}=a$,  then the Beltrami conditions  (\ref%
{BEE}-\ref{BEC}) can be written as
\begin{eqnarray}
\mathbf{\nabla }\times G\mathbf{V}_{h}-\mathbf{B} &=&aG\mathbf{V}_{h}, \\
\mathbf{\nabla }\times G\mathbf{V}_{p}+\mathbf{B} &=&aG\mathbf{V}_{p}, \\
\mathbf{\nabla }\times \mathbf{V}_{c}-\mathbf{B} &=&a_{c}\mathbf{V}_{c}
\mathbf{.}
\end{eqnarray} 
By solving these modified Beltrami conditions along with Ampere's law  (\ref{AL}), one can
derive the following TB equation
\begin{equation}
\left(\mathbf{\nabla }\times\right)^{3}\mathbf{B}-k_{3}\left(\mathbf{\nabla }\times\right)^{2}\mathbf{B}+k_{2}\mathbf{\nabla }\times \mathbf{B}
-k_{1}\mathbf{B=}0,\label{TBS}
\end{equation}
where $k_{3}=a+a_{c}$, $k_{2}=N_{c}+aa_{c}+\left( 1+N_{p}\right)G^{-1}$ and $
k_{1}=aN_{c}+\left( 1+N_{p}\right)G^{-1}$. So the eigenvalues of the TB  equation (\ref{TBS}) can be calculated from the following cubic equation 
\begin{equation}
    \lambda^{3}-k_{3}\lambda^{2}+k_{2}\lambda-k_{1}=0.\label{TBC}
\end{equation}
 Also, the analytical solution of the TB equation (\ref{TBS}) in simple slab geometry is
\begin{equation}
\mathbf{B=}\sum\limits_{j=1}^{3}C_{j}\left[ \sin \left( \lambda _{j}x\right) 
\widehat{y}+\cos \left( \lambda _{j}x\right) \widehat{z}\right], \label{TBA}
\end{equation}
where the values of $\lambda_{j}$ are the roots of equation (\ref{TBC})  and $C_{j}$  are some constants. For the TB state to calculate $C_{1}$, $C_{2}$ and $C_{3}$, we use following boundary
conditions: $\left\vert \mathbf{B}_{z}\right\vert
_{x=0}=\sum\limits_{j=1}^{3}C_{j}=b_{1}$, $\left\vert \mathbf{B}%
_{y}\right\vert _{x=x_{0}}=\sum\limits_{j=1}^{3}C_{j}\sin \left( \lambda
_{j}x_{0}\right) =b_{2}$ and $\left\vert (\mathbf{\nabla \times B}%
)_{z}\right\vert _{x=0}=\sum\limits_{j=1}^{3}C_{j}\lambda _{j}=b_{3}$, where  $b_{1}$, $b_{2}$, and $b_{3}$ are some arbitrary real valued constants.
Now, in order to highlight the impact of cold species density and relativistic temperature on TB magnetic structures for given value of Beltrami parameters figure (\ref{fig:4})  is plotted by using the following values of boundary conditions:  $b_{1}=1.0$, $b_{2}=0.1$ and $b_{3}=0.01$. In figure (\ref{fig:4a}) for $a=3.5$, $a_{c}=0.85$ and $G=4.2$, when $N_{c}=0.001$, the values of all the scale parameters are real ($\lambda_{1}=0.1767$, $\lambda_{2}=0.8063$ and $\lambda_{3}=3.3671$) and consequently  TB relaxed state shows paramagnetic trend, whereas for $N_{c}=0.1$, one of the scale parameter is real and other two are complex conjugate ($\lambda_{1}=3.3686$ and $\lambda_{2,3}=0.4907\pm 0.0669i$) and consequently TB state shows partial diamagnetism.
Similarly the figure (\ref{fig:4b}) shows the impact of relativistic temperature for given value of plasma parameters ($a=0.9$, $a_{c}=3.0$ and $N_{c}=0.01$). So when $G=1.2$, the eigenvalues of TB are $\lambda_{1}=2.3306$ and $\lambda_{2,3}=0.7847\pm 2.3305i$ and plasma shows diamagnetic behavior but in the case of ultrarelativistic regime $G=8.0$, all the eigenvalues are real ($\lambda_{1}=0.1005$, $\lambda_{2}=0.8845$ and $\lambda_{3}=2.9151$) and plasma shows a paramagnetic trend.
\subsection{DB state} \label{ss4.2}

\begin{figure}
     \centering
     \begin{subfigure}[b]{0.49\textwidth}
         \centering
         \includegraphics[width=\textwidth]{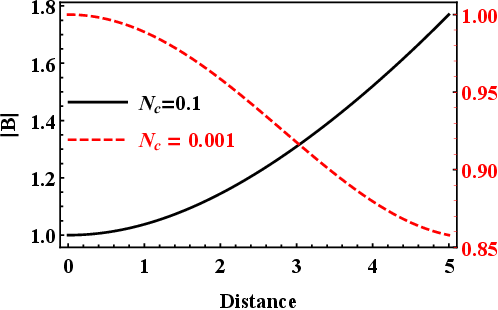}
         \caption{$a=1.5$, $G=4.2$ and $N_{c}=0.1$--(left vertical axis) and $0.001$--(right vertical axis).}
         \label{fig:5a}
     \end{subfigure}
     \hfill
     \begin{subfigure}[b]{0.49\textwidth}
         \centering
         \includegraphics[width=\textwidth]{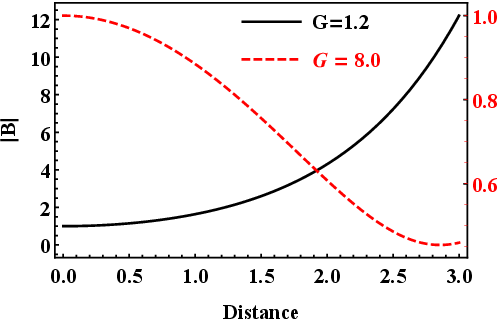}
         \caption{$a=1.5$, $N_{c}=0.01$ and $G=1.2$--(left vertical axis) and $8.0$--(right vertical axis).}
         \label{fig:5b}
     \end{subfigure}
        \caption{Trend of DB magnetic structures for different values of plasma parameters.}
        \label{fig:5}
\end{figure}

Consider another scenario when the flows of different
plasma species are adjusted so that the ratios of generalized vorticities to
flows are the same for all plasma species. In this case all the Beltrami conditions are defined by a
single Beltrami parameter ($a_{h}=a_{p}=a_{c}=a$). Under this assumption, the
Beltrami conditions (\ref{BEE}-\ref{BEC}) can be expressed as
\begin{eqnarray}
\mathbf{\nabla }\times G\mathbf{V}_{h}-\mathbf{B} &=&aG\mathbf{V}_{h}, \\
\mathbf{\nabla }\times G\mathbf{V}_{p}+\mathbf{B} &=&aG\mathbf{V}_{p}, \\
\mathbf{\nabla }\times \mathbf{V}_{c}-\mathbf{B} &=&a\mathbf{V}_{c}\mathbf{.}
\end{eqnarray}
By solving these Beltrami conditions along with equation (\ref{AL}), the following DB equation is obtained
\begin{equation}
 \left(\mathbf{\nabla }\times\right)^{2}\mathbf{B}-k_{2}\mathbf{\nabla 
}\times \mathbf{B}+k_{1}\mathbf{B}=0,  \label{DBS} 
\end{equation}
where $k_{1}=N_{c}+\left( 1+N_{p}\right)G^{-1}$ and $k_{2}=a$. The eigenvalues ($\lambda _{1,2}$) for this DB state (\ref{DBS}) are given by the following quadratic equation
\begin{equation}
    \lambda^{2}-k_{2}\lambda+k_{1}=0.\label{DBC}
\end{equation}
whereas the analytical solution of DB state is
\begin{equation}
 \mathbf{B}=\sum\limits_{j=1}^{2}C_{j}\left[ \sin \left( \lambda _{j}x\right) 
\widehat{y}+\cos \left( \lambda _{j}x\right) \widehat{z}\right],\label{DBA}   
\end{equation}
where $C_{j}$'s  are constants and in order to find $C_{1}$ and $C_{2}$, we use following boundary
conditions: $\left\vert \mathbf{B}_{z}\right\vert
_{x=0}=\sum\limits_{j=1}^{2}C_{j}=b_{1}$and $\left\vert (\mathbf{\nabla
\times B})_{z}\right\vert _{x=0}=\sum\limits_{j=1}^{2}C_{j}\lambda _{j}=b_{2}
$. For the field profiles of the DB state $b_{1}=1.0$ and $b_{2}=0.5$ will be used. Now, figure (\ref{fig:5}) is plotted to show a glimpse of DB magnetic structure and the impact of plasma parameters ($N_{c}$ and $G$) on them for the given value of Beltrami parameter ($a=1.5$). From figure (\ref{fig:5a}) it is very clear for $G=4.2$, when $N_{c}=0.001$ the scale parameters have real values ($\lambda_{1}=0.4580$ and $\lambda_{2}=1.0421$) and relaxed state shows paramagnetic behavior whereas for $N_{c}=0.1$, the scale parameters are complex conjugate ($\lambda_{1,2}=0.75\pm 0.1171i$) and plasma shows diamagnetic structure. Similarly from figure (\ref{fig:5b}) it is very clear for $N_{c}=0.01$, when $G=1.2$, the scale parameters are complex conjugate ($\lambda_{1,2}=0.75\pm 1.0555i$) and plasma shows diamagnetic structure.
Whereas for $G=8.0$, the scale parameters have real values ($\lambda_{1}=0.2$ and $\lambda_{2}=1.3$) and relaxed state shows paramagnetic behavior.

\subsection{Classical perfect diamagnetic state} \label{ss4.3}

 In a study by Mahajan, it was shown that for a single fluid classical plasma, the vanishing of generalized helicity leads to a perfect diamagnetic relaxed state equation similar to London's equation of superconductivity \cite{Mahajan2008}. Mahajan termed this perfect diamagnetic state in classical plasmas as classical perfect diamagnetism . The classical perfect diamagnetic states in plasmas around the black holes are also explored by Bhattacharjee et al \cite{Bhattacharjee2019,Bhattacharjee2020}. Later on Asenjo and Mahajan extended this model to multispecies plasmas. For instance, in the radiation epoch of the early universe, they examined perfect diamagnetic states in an electron-positron plasma. In these expanding cosmic plasmas in curved space-time, the electromagnetic, kinematic, and thermal forces can balance, eliminating the generalized helicities of plasma species and causing a perfect diamagnetic field trend \cite{Asenjo2019}. 
In order to explore the possibility of classical perfect diamagnetism in our plasma model, we now consider that the generalized vorticity or helicities of all the plasma species vanishes
($a_{h}=a_{p}=a_{c}=0$).  It is important to emphasize that the vanishing of generalized helicities of all the plasma species is the singular limit for this plasma model.  In realistic plasmas by injecting the particle beams, the extremely small values of generalized  helicities can be attained.  
Now, for zero generalized helicities the  Beltrami conditions (\ref{BEE}-\ref{BEC}) can be expressed as
by 
\begin{eqnarray}
\mathbf{\nabla }\times G\mathbf{V}_{h}-\mathbf{B} &=&0, \\
\mathbf{\nabla }\times G\mathbf{V}_{p}+\mathbf{B} &=&0, \\
\mathbf{\nabla }\times \mathbf{V}_{c}-\mathbf{B} &=&0,
\end{eqnarray}
which when coupled with equation (\ref{AL}), yields the following relaxed state equation
\begin{equation}
  \left(\mathbf{\nabla }\times\right)^{2}\mathbf{B}+k\mathbf{B}=0,  \label{CPD}
\end{equation}
where $k=N_{c}+\left( 1+N_{p}\right)G^{-1}$. The characteristic equation of this relaxed state is
\begin{equation}
    \lambda^{2}+k=0,\label{CPDC}
\end{equation}
whose roots are $\lambda_{1,2}=\pm i\sqrt{N_{c}+\left( 1+N_{p}\right)G^{-1}}$. Now, we show that for the classical perfect diamagnetic state the cold species density and relativistic temperature can  \textcolor{red}{affect} the strength of magnetic field. For the analytical solution of equation (\ref{CPD}), we use the solution given by equation (\ref{DBA}) with same boundary conditions.
\begin{figure}
     \centering
     \begin{subfigure}[b]{0.49\textwidth}
         \centering
         \includegraphics[width=\textwidth]{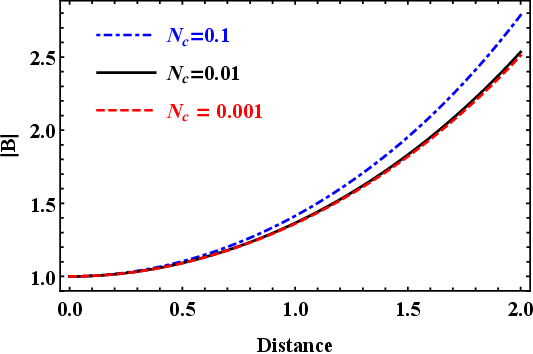}
         \caption{$G=4.2$ and $N_{c}=0.1,0.01$ and $0.001$.}
         \label{fig:6a}
     \end{subfigure}
     \hfill
     \begin{subfigure}[b]{0.49\textwidth}
         \centering
         \includegraphics[width=\textwidth]{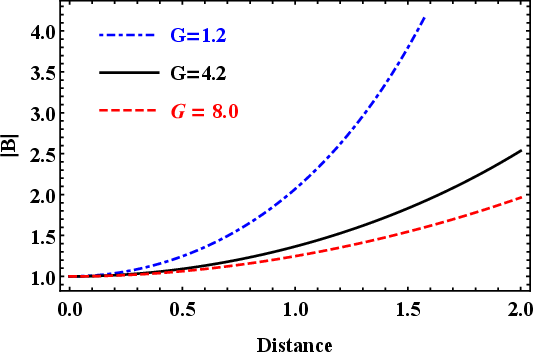}
         \caption{$N_{c}=0.01$ and $G=1.2,4.2$ and $8.0$.}
         \label{fig:6b}
     \end{subfigure}
        \caption{Effect of different values of $N_{c}$ and $G$ on classical perfect diamagnetic structures.}
        \label{fig:6}
\end{figure}
As the eigenvalues of the relaxed state are imaginary so the sine and cosine functions in analytical solution transform to hyperbolic functions which always grow away from the center. This trend is illustrated in figure (\ref{fig:6})  which also shows the impact of cold species density and relativistic temperature on the classical perfect diamagnetic state. From figure (\ref{fig:6a}), shows trend of magnetic field for different values of $N_{c}$ for $G=4.2$.  The eigenvalues for $N_{c}=0.1,0.01$and $0.001$ are $\lambda_{1,2}=\pm 0.7591i, \pm 0.6973i$ and $\pm 0.6908i$, respectively. So, from plot it is evident that by increasing $N_{c}$ , the strength of the magnetic field also increases. This increase in the strength of magnetic field is consequence of increase in the values of scale parameters. On the other hand figure (\ref{fig:6b}) shows the effect of $G$ for $N_{c}=0.01$. The values of scale parameters are $\lambda_{1,2}=\pm 1.2949i, \pm 0.6973i$ and $\pm 0.51i$ for $G=1.2,4.2$ and $8.0$, respectively. The plot clearly shows that as the values of the scale parameters decrease with increasing relativistic temperature, the strength of the magnetic field also decreases.

According to the above discussion, the TB, DB and perfect diamagnetic states can be obtained in this plasma system by adjusting the Beltrami parameters. In addition, it is abundantly obvious, even in the TB and DB states, the formation of the para- and diamagnetic structures can be controlled by cold species density and relativistic temperature, whereas in the perfect diamagnetic state, the strength of diamagnetism also highly depends on these two plasma parameters.

\section{Summary and Conclusion} \label{S6}
The relaxed state of a two electron-temperature relativistic hot EPI plasma has been investigated. The inertial effects of relativistic hot pair species and cold electron species are taken into account, while the heavy positively charged ions are assumed to be static and play their role in quasineutrality.
Starting from the macroscopic evolution equations of inertial plasma species, the vortex dynamics equations are obtained. The steady state solution of vortex dynamics equations provides three Beltrami conditions. When these Beltrami conditions are solved simultaneously, along with Ampere's law, a QB relaxed state is obtained. In the QB state the fields and flows of plasma species also show strong magnetofluid coupling. The eigenvalues of this QB state can be either real or both real and complex. The analysis of QB state shows that at higher relativistic temperatures and lower cold species densities, the eigenvalues become real for the given values of Beltrami parameters. Additionally, the analytical solution of the relaxed state is presented in a simple slab geometry. It has been shown that at lower relativistic temperatures and higher cold species densities, paramagnetic structures transform into diamagnetic ones. Furthermore, it has been shown that when the generalized helicities of hot pair species are equal, the relaxed state is TB state, while for the same generalized helicities of hot pair and cold electron species, the relaxed state is DB state. For TB and DB states, it is also demonstrated that the nature of relaxed field structures can also be transformed by varying cold species density and relativistic temperature for the fixed values of generalized helicities. In the case where the generalized helicities of plasma species vanish, the relaxed state equation is London's equation of superconductivity. This relaxed state shows perfect diamagnetic behavior, and the strength of diamagnetism can be controlled by the cold species density and relativistic temperature of plasma species. It is crucial to emphasize that the conversions between diamagnetic structures and paramagnetic structures are significant within the framework of converting magnetic energy into kinetic energy. Moreover, the multi-Beltrami relaxed states, namely QB, TB, and DB, offer the potential for the formation of multiscale structures. So, the present findings ought to contribute in improving the knowledge of laboratory relativistic hot plasmas containing some cold plasma species as well as those found in cosmic environments such as around pulsar magnetospheres and AGN.

\section*{Compliance with Ethical Standards}
Not applicable
\begin{itemize}
\item \textbf{Funding:} The work of M. Iqbal is supported by Higher Education Commission (HEC), Pakistan, under project No. 20-9408/Punjab/NRPU/R\&D/HEC/2017-18.
\item \textbf{Disclosure of potential conflicts of interest:} No conflicts of interests or competing interests are related to the present work.
\item \textbf{Ethical Approval:} Not applicable.
\item \textbf{Data Availability Statements:} Data sharing is not applicable to this article as no new data were created or analyzed in this study.
\end{itemize}

\end{document}